# Electric-field control of domain wall nucleation and pinning in a metallic ferromagnet

A. Bernand-Mantel,[1,a] L. Herrera-Diez,[1] L. Ranno,[1] S. Pizzini,[1] J. Vogel,[1] D. Givord,[1] S. Auffret,[2] O. Boulle,[2] I. M. Miron[2] and G. Gaudin[2]

[1]Institut Néel, CNRS and Université Joseph Fourier, BP 166, F-38042 Grenoble Cedex 9, France

[2]SPINTEC, UMR-8191, CEA/CNRS/UJF/GINP, INAC, F-38054 Grenoble, France

a)Author to whom correspondence should be addressed. Electronic addresse: anne.bernand-mantel@grenoble.cnrs.fr

The electric (E) field control of magnetic properties opens the prospects of an alternative to magnetic field or electric current activation to control magnetization. Multilayers with perpendicular magnetic anisotropy (PMA) have proven to be particularly sensitive to the influence of an E-field due to the interfacial origin of their anisotropy. In these systems, E-field effects have been recently applied to assist magnetization switching and control domain wall (DW) velocity. Here we report on two new applications of the E-field in a similar material : controlling DW nucleation and stopping DW propagation at the edge of the electrode.

The possibility of controlling magnetic properties with an E-field via magneto-electric (ME) effects was initially envisaged in multiferroïcs[1] or magnetic semiconductors[2]. As the majority of these materials loses their ferromagnetic properties at room temperature the recent discovery of an important ME effect at room temperature in conventional ferromagnetic metals[3-8] has attracted much attention. Despite the limited penetration depth of an E-field in metals, the charge induced at the metal/dielectric interface on the topmost atomic layers is sufficient to modify the surface magnetic anisotropy energy (MAE) by a non negligible amount in ultrathin ferromagnetic layers as suggested by electronic structure calculations[9-11]. This has a particular impact in systems where the different anisotropy contributions almost cancel each other out, resulting in a large relative variation of the total effective MAE when the surface contribution is modified with the voltage[12-26]. We have studied here a Pt/Co/AlOx sample where the surface MAE could be varied in two ways : charging the metal/dielectric interface and modifying its oxidation. We demonstrate that charging and oxidizing the interface affect the magnetic properties in some equivalent manner in this system. The E-field effect is characterized by monitoring the magnetization reversal for different MAE values. For weak MAE, we observe a nucleation-dominated reversal and we demonstrate that the thermally activated DW nucleation can be electrically controlled via the modulation of the involved energy barrier. For higher MAE we observe a strong E-field variation of the thermally activated DW creep motion, as reported in previous works[22-24]. In addition to this E-field control of DW velocity below the electrode we demonstrate the reversible pinning of DW at the edge of the electrode. The measurements reported here are carried out using magneto-optical Kerr microscopy. The E-field is applied locally via transparent electrodes deposited on the un-patterned dielectric/metalic bilayer, thus allowing direct comparison of the magnetic properties with and without the E-field within the same image. The metallic ferromagnetic multilayer consists of a Si/Pt(3 nm)/Co(0.6 nm)/Al(wedge ~1.5 nm) stack prepared by sputtering. The Al wedge is oxidized in a $O_2$ plasma, resulting in the formation of an $AlO_X$ layer. The 0.1 nm for 1 cm linear increase in the Al thickness along the wedge results in a progressive variation of the Co/AlOx interface oxidation. This oxidation has a direct impact on the magnetic anisotropy of the Co layer : for an optimum oxidation of the Co/AlOx interface the ferromagnetic layer possesses an out of plane anisotropy[27]. The alumina layer is subsequently covered by an Atomic Layer Deposition (ALD) grown $Al_2O_3$(~7nm)/$HfO_2$ (~37nm) dielectric bilayer. Transparent top gate electrodes of 50 x 700 μm wide indium tin oxide (ITO) are deposited on top by sputtering and patterned by UV lithography. A schematic representation of the device is showed in Figure 1.

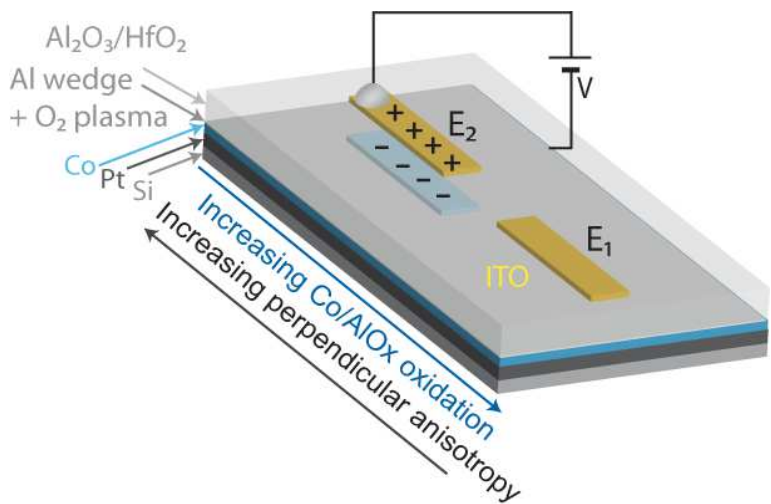

**Fig. 1:** Schematic representation of the magnetic multilayers and the device structure. The direction of increasing perpendicular anisotropy and increasing Co/AlOx oxidation are indicated. The two electrodes under which the E-field effect have been characterized in the low ($E_1$) and high ($E_2$) anisotropy regions are represented.

To study the E-field influence on DW nucleation we have selected a low anisotropy (higher $Co/AlO_X$ oxidation) region compared to the high anisotropy (optimum oxidation) region.

In this region a high nucleation density was observed at the time, magnetic field and length scales of our experiment. A Kerr image taken below the ITO electrode $E_1$ 200 ms after applying a constant 1 mT magnetic field is shown in Figure 2(a) (left).

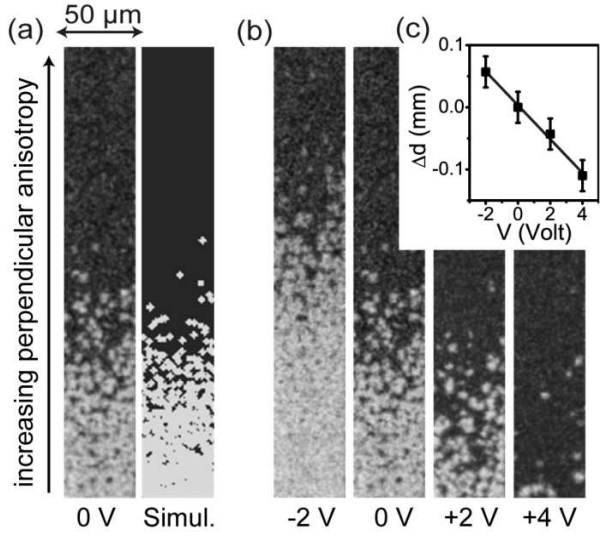

**Fig. 2:** (a) (left) and (b) Kerr images recorded through the ITO electrode $E_2$ 200 ms after a 1 mT constant magnetic field is applied and for different applied voltages. (a) (right) Simulation of the nucleation. (c) Displacement of the 50% reversed zone as a function of the applied voltage.

Regions where the magnetization is unreversed (reversed) appear in dark grey (light grey). The nucleation density is increasing (indicating a decrease in MAE) in the direction where the Co/AlOx interface is getting more oxidized. If we refer to previous works on this system[27], this confirms that the Co/AlOx interface is in the over-oxidized state in our sample. The fast variation of nucleation density suggests that the nucleation rate follows some exponential law as expected for a thermally activated nucleation process above an energy barrier. To confirm this we have calculated the nucleation probability within the Néel-Brown model (see Appendix A ). The results are presented in Figure 2.a (right). The calculation provides a qualitative description of the experimental data for a linear variation of the energy barrier height along the wedge $\Delta E_N / \Delta x = 0.22$ meV/μm and an energy barrier height $E_N = 500$ meV (at the bottom on the image in Figure 2(a)). The influence of the E-field on the nucleation is checked by comparing Kerr images taken 200 ms after applying a 1 mT magnetic field for various V (Figure 2(b)). The region where 50% of the magnetization is reversed is linearly shifted by the E-field with a position/voltage equivalence of $\Delta x / \Delta V = 30$ μm/V. This shift toward lower MAE regions for positive voltage is consistent with the MAE increase upon electron density increase already observed in previous works[22,24]. In addition to this, we demonstrate here an equivalence between a reduction of the electron density and an oxidation increase, in the case of an over oxidized Co/AlOx interface and we show that this E-field induced MAE variation can cause an exponential variation of the nucleation rate.

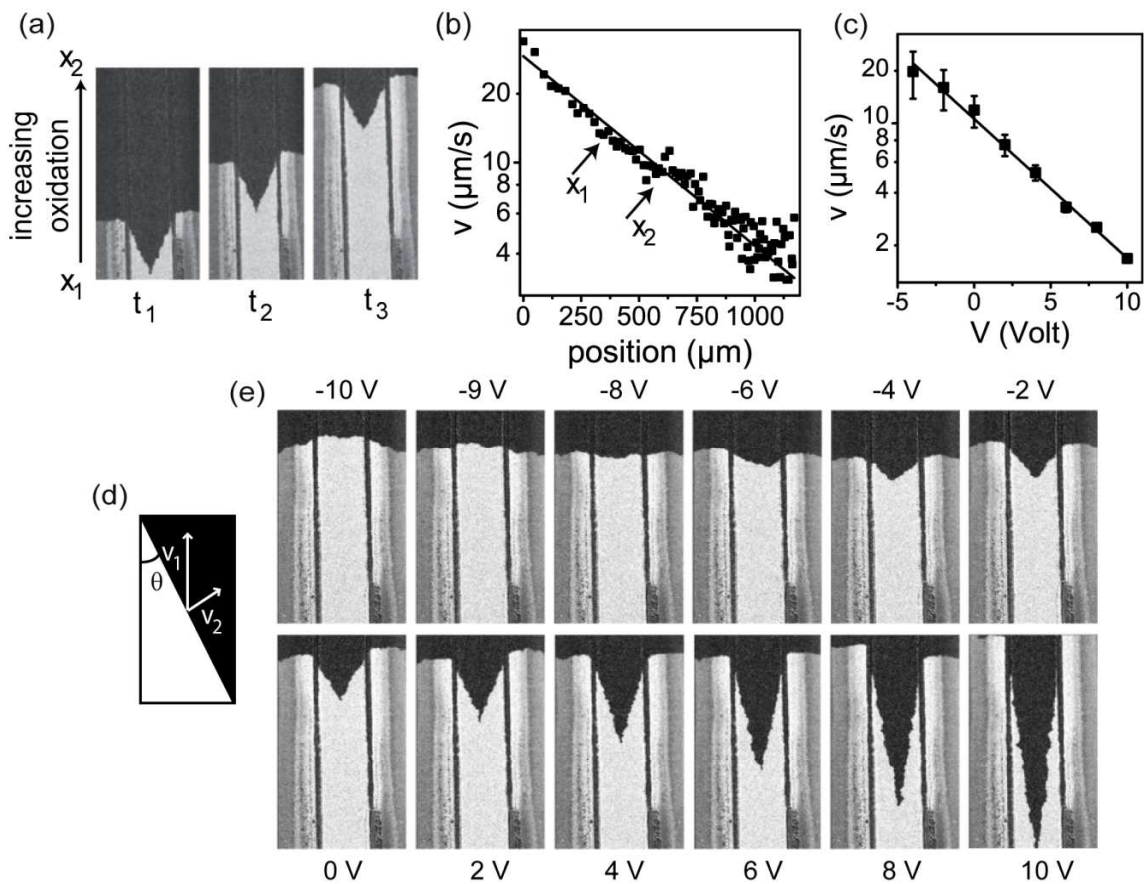

**Fig. 3:** (a) and (e) Kerr images of the magnetization reversal below the ITO electrode (centre) recorded **a,** 12, 16 and 22 s after a constant 2.4 mT magnetic field is applied and (e) for various V. The edges of the electrode are visible as two black vertical lines. (b) and (c) The black squares represent the experimental DW velocities measured (c) outside the ITO electrode at different positions on the wedge (b) below the ITO electrode around position $x_1$ for different applied V. (b) and (c) The black lines represent a fit with an exponential decay function $v = v_0 \exp(-\alpha_{1,2} x)$

where $\alpha_{1,2}$ correspond to the coefficient found for respectively (b) and (c). The ratio $\alpha_2/\alpha_1$ gives the position/voltage equivalence $\Delta x/\Delta V=80$ µm/V. $x_1$ and $x_2$ positions are indicated in (a). (c) DW velocities measured. (a) and (c) The black line represent (d) Definition of the DW velocities outside ($v_1$) and below ($v_2$) the electrode and the angle θ between the DW and the electrode side.

The voltage influence on DW propagation was investigated in the higher anisotropy (less oxidized) region (electrode $E_2$). In this region, for sufficiently small magnetic field, the reversal is dominated by the propagation of a DW created in the lower anisotropy side of the sample where the nucleation field is lower. The DW velocity outside the electrode (i.e. in the absence of E-field) as a function of the position on the wedge for a fixed magnetic field of 2.4 mT is presented in Figure 3(b). The DW velocity decreases due to the increase in MAE along the wedge. The exponential decrease found indicates that the DW motion is in a creep regime i.e. the wall propagates by thermal activation over energy barriers determined by local anisotropy fluctuations. This mechanism is consistent with the irregular aspect of the DW. A similar analysis has been proposed in previous studies on similar materials[22-24,30]. The E-field influence on DW propagation is obtained by examining DW propagation below the electrode. The DW presents a triangular shape which remains unmodified while propagating (Figure 3(a)). The presence of this triangle suggest a difference in the DW velocity outside the electrode, and below the electrode where the film is submitted to the E-field. The triangle propagates along the vertical direction at the same velocity $v_1$ as the DW outside the electrode. However the real DW velocity below the electrode is represented by the motion of the triangle sides. It is equal to $v_2$, the projection of $v_1$ along the direction perpendicular to the triangle side. $v_2/v_1=\sin\theta$ where θ is the angle between the side of the triangle and the side of the electrode (see Figure 3.d). The angle θ is decreasing with increasing V, illustrating the reduction in the DW velocity occurring under a positive applied voltage (Figure 3.e). Note that a flat domain wall ($v_1 = v_2$), expected for V=0 is obtained for a voltage of -8 V. This effect may be attributed to the presence of fixed charges inside the dielectric which have to be compensated to reach the effective zero E-field (see Appendix B). The DW velocity below the electrode, extracted from each Kerr image at position $x_1$ varies exponentially with V (Figure 3(c)). This indicates, as reported in previous works, that the DW pinning energy barrier is modulated by the E-field via the MAE variation[22-24]. By comparing Figure 3 b and c we see that applying a negative voltage has the same effect on the DW velocity as a displacement toward a more oxidized region. Consequently, decreasing the electron density has the same effect on the MAE as increasing the interface oxidation as observed in the lower anisotropy region (below $E_1$). The position/voltage equivalence of $\Delta x/\Delta V=80$ µm/V found below $E_2$ is 2.4 times larger than the one found below $E_1$. We can deduce that the effect on MAE of the E-field ($\Delta MAE/\Delta V$) relatively to the effect of oxidation ($\Delta MAE/\Delta x$), is larger in the region with higher MAE.

The use of local electrodes on a continuous ferromagnetic layer allows the creation of an MAE step at the edge of the electrode when a voltage is applied. Such energy step may be used to pin DW. Indeed, we observe that the effect of the E-field on DW propagation differs below and at the edges of electrode $E_2$. In Figure 4, the magnetization reversal at a fixed magnetic field of 2.4 mT is recorded as a function of time and while turning the E-field on and off. The integrated Kerr intensity below the electrode and the applied potential are represented as function of time in Figure 4(a) and the corresponding images are shown in Figure 4(b) (see supplementary material for a video).

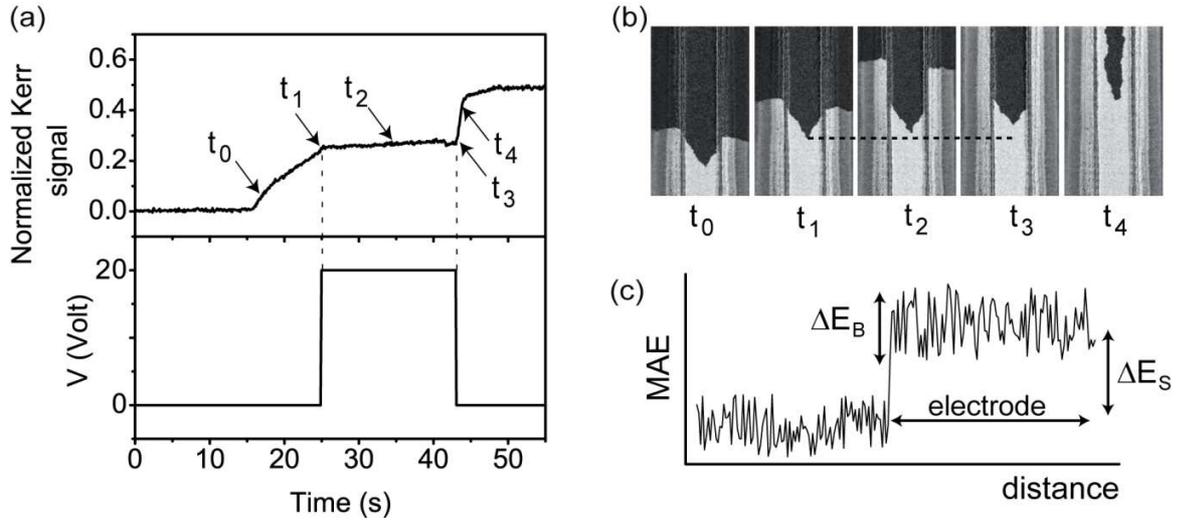

**Fig. 4:** (a) Integrated Kerr intensity recorded below the ITO electrode and applied voltage as a function of time for a constant magnetic field of 2.4 mT. (b) Corresponding Kerr images. (c) Schematic representation of the MAE profile below and outside the electrode. The energy step $\Delta E_S$ and the MAE fluctuations due to local defects $\Delta E_B$ are represented.

DW propagation below the electrode at $t_0$, is characterized by a progressive increase in the detected integrated intensity. At $t_1$ the voltage is turned to +20 V and as a result of the increase in the MAE, the DW velocity decreases by approximately one order of magnitude below the electrode while the DW outside the electrode keeps propagating (Figure 4(b) $t_1$ to $t_3$). Although the DW below the electrode is only slowed down by the E-field, the DW is pinned at the electrode sides between $t_1$ and $t_3$. We argue that this pinning of the DW is related to the presence at the edge of the electrode of an energy step $\Delta E_S$ due to the voltage induced MAE difference below and outside the electrode (see Figure 4(c)). This energy step can be overcome by thermal fluctuations and as its height is proportional to the E-field, an exponential variation of the pinning time with V is expected. We observe that for V=20 V, the DW is efficiently pinned on the electrode edge at the time scale of our experiment (t~20s). At $t_3$, the MAE is lowered and the decrease in the anisotropy step height allows DWs to propagate from both sides of the electrode (last image in Figure 4(c)). This demonstration of an efficient and reversible electrical DW pinning open the way for controlled DW propagation which is essential in future logic or memory applications.

In summary, the results presented here provide a consistent picture of domain wall control in both the nucleation and the propagation regimes using an E-field. The tuning of anisotropy along the wedge allows us to use different local gates on a single sample to either favour DW nucleation below the gate or pin a DW on its edge. Such tuning of the magnetization reversal suggests that magnetic reservoirs and notches, which are respectively used to generate and block DW in nanowires of similar materials, could be advantageously replaced by local gates taking advantage of the tunability of an electric control.

This work was supported by the French National Research Agency (ANR) under the project ANR-2010-BLANC-1006-ELECMADE and the Nanosciences Foundation under the project POMME. The lithography was carried out in the Nanofab cleanroom facilities of Grenoble. The authors thank E. Wagner for technical assistance on the Kerr microscope and M. Darques for his help on image processing.

APPENDIX A : Nucleation simulation

The nucleation probability is defined, following the Néel-Brown model as $P(t) = \left(1 - e^{-t/\tau}\right)$ where $1/\tau = 1/\tau_0 . e^{(-E_N/k_B T)}$ is the nucleation rate, $1/\tau_0 = 10^{10}$ the try rate, $E_N$ the energy barrier and T=300 K the temperature. The image presented in Figure 2(b) right is obtained by a Monte-Carlo simulation on a rectangle where $E_N/k_B T$ is increasing linearly from bottom to top with $E_N/k_B T = 20$ (bottom) and $\Delta E_N/E_N = 12\%$. The nucleation process was simulated for a t=1s characteristic time. To compare the experimental and simulated images we have compared the slope of the averaged horizontal intensity versus vertical position in the image. In the Kerr image of Figure 2.a left, the nucleation centers present a few micron diameter because the DW have started to propagate. To reproduce these experimental conditions we have also propagated the nucleated centers in the simulation.

APPENDIX B : Gate properties and trapped charges.

The measured capacitances of the electrodes are C~0.06 µF/cm$^2$. This agrees with an effective dielectric constant $\varepsilon_{eff}$~15 for the $Al_2O_3/HfO_2$ dielectric bilayer. As observed in Figure 4.e the effective zero E-field is obtained for V~-8 V. This indicates the presence of fixed trapped charges in the dielectric. The derived charge density is of 1.5x10$^{13}$e-/cm². Those charges are induced during the deposition of ITO as they are present only below the ITO electrodes inducing a difference in the anisotropy below ITO and outside ITO. Such trapped charges, with the same density, were observed below ITO electrodes all over the sample. Charges traps with similar densities have already been observed in high-k based capacitors containing $HfO_2$[28] and $ZrO_2$[29] insulators.

REFERENCES

[1]R. Ramesh and N. A. Spaldin, Multiferroics: progress and prospects in thin films, *Nature Mater.* **6**, 21 (2007).

[2]H. Ohno, D. Chiba, F. Matsukura, T. Omiya, E. Abe, T. Dietl, Y. Ohno, K. Ohtani, Electric-field control of ferromagnetism. *Nature* **408**, 944 (2000).

[3]M. Weisheit, S. Fähler, A. Marty, Y. Souche, C. Poinsignon and D. Givord, Electric field-induced modification of magnetism in thin-film ferromagnets *Science* **315**, 349 (2007).